\documentclass[12pt]{article}
\usepackage{graphicx}

\usepackage{mathptmx}
\usepackage{amsmath,amssymb}

\usepackage{times}
\usepackage[T1]{fontenc}
\normalfont
\usepackage{mathptm}
\usepackage{fourier} 

\begin{document}
\renewcommand{\thefootnote}{\#\arabic{footnote}} 
\setcounter{footnote}{0}

\begin{center}

\begin{flushright}
OU-HET 973
\end{flushright}

\vspace{1cm}

\begin{Large}
{\bf Tomography by neutrino pair beam }
\end{Large}

\vspace{1cm}
Takehiko Asaka, Hisashi Okui$^1$,  Minoru Tanaka$^2$ and
Motohiko Yoshimura$^3$

\vspace{0.5cm}

{\it Department of Physics, Niigata University, 950-2181 Niigata, Japan}
\\[2ex]

$^1$
{\it Graduate School of Science and Technology, Niigata University, 950-2181 Niigata, Japan}
\\[2ex]

$^2$
{\it Department of Physics, Graduate School of Science, \\
             Osaka University, Toyonaka, Osaka 560-0043, Japan}
 \\[2ex]

 $^3$
{\it Research Institute for Interdisciplinary Science, Okayama University,\\
Tsushima-naka 3-1-1 Kita-ku Okayama
700-8530 Japan}
\end{center}

\begin{center}
  (May 28, 2018)
\end{center}

\vspace{0.5cm}

\begin{abstract}
We consider tomography of the Earth's interior using the neutrino
pair beam which has recently been proposed.  The beam produces a large
amount of neutrino and antineutrino pairs from the circulating
partially stripped ions and provides the possibility to measure
precisely the energy spectrum of neutrino oscillation probability
together with a sufficiently large detector.  It is shown that the
pair beam gives a better sensitivity to probe the Earth's crust
compared with the neutrino sources at present.  In addition we present
a method to reconstruct a matter density profile by means of the analytic
formula of the oscillation probability in which the matter effect is
included perturbatively to the second order.
\end{abstract}

\vspace{4cm}
\hspace{0.5cm} 

\newpage
\section{Introduction}
Our understanding of neutrino has improved greatly since the end of
the last century.  The observation of flavor oscillations of neutrinos
has shown the presence of non-zero masses of neutrinos contrary to the
prediction of the Standard Model.  This is a clear signature of new
physics beyond the Standard Model.  Thanks to the remarkable efforts
of various neutrino experiments, the mass squared differences and
mixing angles of (active) neutrinos have been measured very precisely
at present ~\cite{Esteban:2016qun}.  The origin of neutrino masses is,
however, unknown and then it is important to investigate the
fundamental theory of neutrino physics.  Moreover, it should be useful
to consider seriously the application of neutrino physics to various
fields of basic science.

In the Standard Model neutrinos are unique matter particles which
possess only the weak interaction (in addition to gravity), and their
interaction rates are very suppressed accordingly.  It is found that
most of them can penetrate the Earth without a scattering when the
energy is smaller than ${\cal O}(10^5)$~GeV~\cite{Giunti_Kim}.  This
shows that neutrinos can be used to probe the deep interior of the
Earth.

The idea of the neutrino tomography has been pointed out in the
1970s~\cite{Placci_Zavattini:1973,Volkova_Zatepin:1974}.  By measuring
the absorption rates of neutrinos passing through the object from
different angles, the image of the object can be reconstructed.  This
is similar to the computed tomography using x-rays, which enables
us to probe inside solids without destruction.  This method is called
as the the neutrino absorption
tomography~\cite{Placci_Zavattini:1973}--\cite{Donini:2018tsg}.  In
addition to this there have been proposed two other methods of
neutrino tomography.  One is the method using neutrino
oscillations~\cite{Ermilova:1988pw}--\cite{Ioannisian:2017chl}, and
the other is the tomography using neutrino
diffraction~\cite{Fortes:2006,Lauter:2017uag}.

The former one utilizes the energy spectrum of the neutrino
oscillation probability, which is distorted, compared to the vacuum
one, by the interaction with matter through which neutrinos pass from
the production to the detection point.  The distortion pattern depends
on the profile of the number density of electron in matter, that can
be translated into the matter density profile by assuming the charge
neutrality and the equality of neutron and proton numbers in matter.
It is then possible to probe the deep interior of the Earth by
measuring the oscillation probability at the sufficient accuracy.

In this letter we revisit the possibility to realize the neutrino
oscillation tomography.  The main difficulties of its feasibility
include the lack of the powerful neutrino source and no established
method to reconstruct the profile of the Earth's interior compared
with the medical computed tomography.  As for the first difficulty we
consider the neutrino pair beam proposed in
Refs.~\cite{Yoshimura:2015rva,Yoshimura:2015ujr}.  It has been shown
that pairs of neutrino and antineutrino can be produced from the
partially stripped ions in circular motion at a larger rate than the
current neutrino sources from pion and muon decays.

On the other hand, the second one is inherent in the tomography using
the oscillation between flavor neutrinos.  It has been
shown~\cite{deGouvea:2000un,Miura:2001pia,Akhmedov:2001kd} that the
flavor oscillation probability with the density profile $\rho (x)$ is
the same as that with $\rho(L-x)$ where $x=0$ or $L$ is the production
or detection position, if only two flavors of neutrinos are
considered.  In general the $\nu_\alpha \to \nu_\beta$ oscillation
probability $P(\nu_\alpha \to \nu_\beta)$ with $\rho(x)$ is equal to
$P(\nu_\beta \to \nu_\alpha)$ with $\rho(L-x)$ and an opposite sign of
the Dirac-type CP-violating phase~\cite{Akhmedov:2001kd}.  Because of
the unitarity conditions
$1 = \sum_\alpha P( \nu_\alpha \to \nu_\beta) =\sum_\beta P(
\nu_\alpha \to \nu_\beta)$ and the absence of the Dirac-type
CP-violation in the two-flavor case $P( \nu_\alpha \to \nu_\beta)$ is
invariant under $\rho(x) \to \rho (L-x)$.  Even for the realistic
three flavor case the invariance holds for the oscillations
$\nu_e \to \nu_e$ and
$\bar \nu_e \to \bar \nu_e$~\cite{Kuo:1987km,Minakata:1999ze} since
they are independent on the Dirac-type phase.%
\footnote{ This discussion cannot be applied to the case with sterile
  neutrinos~\cite{Akhmedov:2001kd}.  } In such cases unambiguous
reconstruction of $\rho(x)$ is possible only if the profile has the
symmetric property with $\rho (x) = \rho (L-x)$.  Otherwise, there
exist degenerate solutions of reconstruction.  It has been, however,
proposed that the difficulty can be avoided by using the transition
probability of mass eigenstate to flavor eigenstate, which can be
realized for the solar and supernova
neutrinos~\cite{Akhmedov:2004rq,Akhmedov:2005yt}.  Here we focus on
the reconstruction of the symmetric density profile with
$\rho (x) = \rho (L-x)$, and provide a useful procedure of its
reconstruction.  Procedures so far proposed are based on the $\chi^2$
analysis (see, for example,
Refs.~\cite{Ohlsson:2001ck,Ohlsson:2001fy}), the inverse Fourier
transformation~\cite{Akhmedov:2005yt}, and so on.  The advantage of
ours is that the reconstruction with a sufficient spatial resolution
is possible even with a low numerical cost.

This letter is organized as follows: In section 2 we briefly review
the neutrino oscillation in matter and present the analytic formula of
the oscillation probability based on the perturbation of the matter
effect, which will be used to reconstruct the density profile
$\rho(x)$.  In section 3 it is discussed the oscillation tomography 
using the neutrino pair beam.  We show the possibility of the
tomography under the ideal situation.  It is then considered how to
reconstruct $\rho(x)$ in section 4.  Finally, our results are
summarized in section 5.

\section{Neutrino Oscillation in Matter}
We begin with briefly reviewing the neutrino oscillation with matter
effects~\cite{Wolfenstein:1977ue,Mikheev:1986gs,Mikheev:1986wj}.  The
transition amplitude of the $\nu_\alpha$ $\to$ $\nu_\beta$ oscillation
($\alpha, \beta = e, \mu, \tau$) at the distance $x$ from neutrino
source is denoted as
\begin{align}
  A_{\beta \alpha}(x) =
  \langle \nu_\beta | \nu_\alpha (x) \rangle \,,
\end{align}
where the initial condition is $| \nu_\alpha (0) \rangle =
| \nu_\alpha \rangle$.
It satisfies the following evolution equation.
\begin{align}
  \label{eq:EQ_A}
  i \frac{d}{dx} A_{\beta \alpha} (x)
  =
  \left[ H_0^F + V^F (x) \right]_{\beta \gamma} \, A_{\gamma \alpha} (x) \,.
\end{align}
Here and hereafter we assume that all the neutrinos are ultrarelativistic.
The free Hamiltonian in the basis of flavor neutrinos is
$H_0^F$ is given by
\begin{align}
  H_0^F &= U \, H_0 \, U^\dagger  ~~~
  \mbox{with}~~ H_0 = \mbox{diag}
  \left(\, \frac{m_1^2}{2 E_\nu} \,,~\frac{m_2^2}{2 E_\nu} \,,~
  \frac{m_3^2}{2 E_\nu} \,\right) \,,
\end{align}
where $E_\nu$ is a neutrino energy, $m_i$ ($i=1,2,3$) is a neutrino mass eigenvalue and
$U_{\alpha i}$ is the Pontecorvo-Maki-Nakagawa-Sakata (PMNS) mixing
matrix~\cite{Pontecorvo:1958,Maki:1962mu}.  The effective potential
in the flavor basis is given by
\begin{align}
  \label{eq:VF}
  V^F (x) &= \mbox{diag} 
        \left( \sqrt{2}\,G_F \, n_e (x) \,,~0\,,~0 \right) \,,
\end{align}
where $G_F$ is the Fermi constant and $n_e(x)$ is the number density of
electrons at the distance $x$.  By taking $n_p = n_n =n_e$ and
$m_p = m_n$ in matter $n_e$ can be written as
\begin{align}
  n_e (x) = \frac{\rho (x) }{2 m_p} \,.
\end{align}
The matter density profile is denoted by $\rho(x)$.
The oscillation probability measured by the detector
at the distance $x=L$ is given by
\begin{align}
  P(\nu_\alpha \to \nu_\beta ; E_\nu) =  \big| A_{\beta \alpha }(L) \big|^2 \,,
\end{align}
where the initial condition of the amplitude is
$A_{\alpha \gamma } (0) = \delta_{\alpha \gamma}$.
On the other hand, the amplitude of the
antineutrino mode
$\bar A_{\beta \alpha}(x) = \langle \bar \nu_\beta | \bar \nu_{\alpha
}(x) \rangle$ is obtained by the replacements
$U \to U^\ast$ and $V^F \to - V^F$. 

The oscillation probability for a given matter density profile can be
obtained by solving numerically Eq.~(\ref{eq:EQ_A}).  On the
other hand, when the matter effect is sufficiently small, it is
obtained analytically based on the perturbation
theory~\cite{Miura:2001pia,Ioannisian:2004jk}.  We shall expand the
amplitude as
\begin{align}
  A_{\beta \alpha }(x) =
  A^{(0)}_{\beta \alpha }(x) +
  A^{(1)}_{\beta \alpha }(x) +
  A^{(2)}_{\beta \alpha }(x) + \cdots \,,
\end{align}
where $A^{(n)}_{\beta \alpha }(x)$ is the $n$-th order correction
of the matter effect.
The explicit expressions
up to the second order are
\begin{align}
  A_{\beta \alpha}^{(0)}(x)
  &= U_{\beta j} U_{\alpha j}^{*} e^{-i E_{j} x}  \,,
  \\
  A_{\beta \alpha}^{(1)}(x)
  &=
    - i U_{\beta j} U_{e j}^{*} U_{e k} U_{\alpha k}^{*} e^{-i E_{j} x}
    \int_{0}^{x} dx_{1} e^{i (E_{j}-E_{k}) x_{1}} v(x_{1}) \,,
  \\
  A_{\beta \alpha}^{(2)}(x)
  &=
    -U_{\beta j} U_{e j}^{*} U_{e k} U_{e k}^{*} U_{e l} U_{\alpha l}^{*}
    e^{-i E_{j} x}
    \int_{0}^{x} dx_{1} \int_{0}^{x_{1}} dx_{2} e^{i (E_{j}-E_{k}) x_{1}}
    e^{i (E_{k}-E_{l}) x_{2}} v(x_{1}) v(x_{2})  \,,
\end{align}
where $E_i = \frac{m_i^2}{2 E_\nu}$ and $v(x)$ is given by the density profile as
\begin{align}
  v(x) = \frac{G_F}{\sqrt{2} m_p} \rho(x) \,.
\end{align}
The oscillation probabilities 
at $x=L$
up to the second order are then given by
\begin{align}
  P^{(0)} (\nu_\alpha \to \nu_\beta ; E_\nu)
  &= | A^{(0)}_{\beta \alpha} (L) |^2 \,,
  \\
  P^{(1)} (\nu_\alpha \to \nu_\beta ; E_\nu)
  &=  A^{(0)}_{\beta \alpha}{}^\ast (L)
    A^{(1)}_{\beta \alpha}{}^\ast (L)
    + h.c. \,,
  \\
  P^{(2)} (\nu_\alpha \to \nu_\beta ; E_\nu)
  &= | A^{(1)}_{\beta \alpha} (L) |^2
    +
    \left[
    A^{(0)}_{\beta \alpha}{}^\ast (L)
    A^{(2)}_{\beta \alpha}{}^\ast (L)
    + h.c.
    \right]
    \,,
\end{align}

When there are only two flavors of neutrinos ($\nu_e$ and $\nu_\mu$),
the mixing matrix is given by
\begin{align}
U=\begin{pmatrix}
\cos{\theta} & \sin{\theta}\\
-\sin{\theta} & \cos{\theta}
\end{pmatrix} \,,
\end{align}
which leads to the zeroth and first order probabilities
\begin{align}
  \label{eq:P0}
  P^{(0)}(\nu_{e} \rightarrow \nu_{e};E_\nu)
  &= 1- \sin^{2} (2\theta) \sin^{2}
    \left( \frac{\Phi L}{2}  \right) \,,
  \\
  \label{eq:P1}
  P^{(1)}(\nu_{e} \rightarrow \nu_{e};E_\nu)
  &=
    \frac{1}{2} \sin^{2} (2\theta ) \cos (2\theta )
    \int_{0}^{L} dx_1 v (x_1)
    \big[ \sin (\Phi L) - \sin (\Phi x_1) - \sin (\Phi (L-x_1))
    \big] \,,
\end{align}
where the mass squared difference $\Delta m^2 = m_2^2 - m_1^2$
and 
\begin{align}
\Phi=E_2-E_1=\frac{\Delta m^{2}}{2E_\nu} \,.
\end{align}
On the other hand, the second order expression is given by
\begin{align}
  P^{(2)} (\nu_{e} \rightarrow \nu_{e}; E_\nu)
  &=
    P^{(2a)}(\nu_{e} \rightarrow \nu_{e}; E_\nu)
    +
    P^{(2b)}(\nu_{e} \rightarrow \nu_{e}; E_\nu) \,,
\end{align}
where
\begin{align}
  P^{(2a)}(\nu_{e} \rightarrow \nu_{e}; E_\nu)
  =
  &
    \big[ \cos^{8}{\theta}+\sin^{8}{\theta}+2\cos^{4}{\theta}
    \sin^{4}{\theta}\cos{(\Phi L)} \big] \, G_{1}^2(L)
    \nonumber \\
	&+
   \cos^{4}{\theta}\sin^{4}{\theta} \,
   \big[ G_{2}^2(L) +G_{3}^2(L) \big]
   \nonumber \\
	&+
   2 \big( \cos^{4}{\theta}+\sin^{4}{\theta} \big)
   \cos^{2}{\theta}\sin^{2}{\theta} \, G_{1}(L) \, G_{2}(L) \,,
  \\
  P^{(2b)}(\nu_{e} \rightarrow \nu_{e}; E_\nu)
  =
  & - 2\int^{L}_{0} dx_{1} \int^{x_{1}}_{0}dx_{2} \,
    v(x_{1}) v(x_{2})  \,
    \bigg\{ + \cos^{8}{\theta}+\sin^{8}{\theta}
    \nonumber \\
  &~~~~~
    + \cos^{2}{\theta}\sin^{2}{\theta} \,
    \big( \cos^{4}{\theta}+\sin^{4}{\theta} \big)
    \big[ \cos{(\Phi L)}+\cos{(\Phi x_{2})}
    \nonumber \\
  &~~~~~~~~~~~~~~~~~~~~~~~~~~~~~~~~~~~~~~~
    +\cos{(\Phi (x_{2} - x_{1}))}
    +\cos{(\Phi (x_{1}-L))} \big]
    \nonumber \\
  &~~~~~
    + 2 \cos^{4}{\theta}\sin^{4}{\theta} \,
    \big[ \cos{(\Phi (x_{2} - L))}+\cos{(\Phi x_{1})}
    +\cos{(\Phi (x_{2} - x_{1}+L))}  \big]
    \bigg\} \,.
\end{align}
Here we have introduced the functions
\begin{align}
  G_{1}(x)
  &= \int_{0}^{x}dx_{1} v(x_{1}) \,,
  \\
  G_{2} (x)
  & = \int_{0}^{x} dx_1 v(x_{1})
    \Big[ \cos \left( \Phi x_{1} \right) +
    \cos \left( \Phi (x-x_{1}) \right) \Big] \,,
  \\
  G_{3} (x)
  & = \int_{0}^{x} dx_1 v (x_{1})
    \Big[ \sin (\Phi x_{1}) + \sin (\Phi (x-x_{1}) ) \Big] \,.
\end{align}
Note that the probabilities for the antineutrino mode
is obtained by replacing $v(x) \to - v(x)$.
The energy spectrum of the oscillation probability
obtained by the perturbation theory will be used
to reconstruct the matter density profile,
which will be discussed in Section~\ref{sec:reconstruction}.

In the present analysis the oscillation with two-flavor
neutrinos ($\nu_e$ and $\nu_\mu$) is investigated for the sake of simplicity. 
The study with the realistic three flavor case will be
discussed elsewhere~\cite{Asaka:2018xxx}.
The mixing angle $\theta$ and mass squared difference $\Delta m^2$ in vacuum
are taken as
\begin{align}
  \sin^2 \theta = 0.306 \,,~~~
  \Delta m^2 = 7.50 \times 10^{-5}~\mbox{eV}^2 \,,
\end{align}
which correspond to those associated with the solar neutrino
oscillation~\cite{Esteban:2016qun}.

We consider for instance the matter
density profile given by
\begin{align}
  \label{eq:rho_ex}
       \rho(x)
       = \bar \rho +
       (\rho_{l} - \bar \rho)
       \exp \left[- \frac{ \left(x- \frac{L}{2} \right)^{2}}{D_{l}^{2}} \right] \,, 
\end{align}
where the background mass density is $\bar \rho$ and
it is taken as $\bar \rho = 2.7$~g/cm$^3$
by considering the continental crust.
In addition the lump with a density $\rho_l$ and a width $D_l$
is located at the center $L/2$.

\begin{figure}[t]
  \centerline{
    \includegraphics[height=8.5cm,pagebox=cropbox]{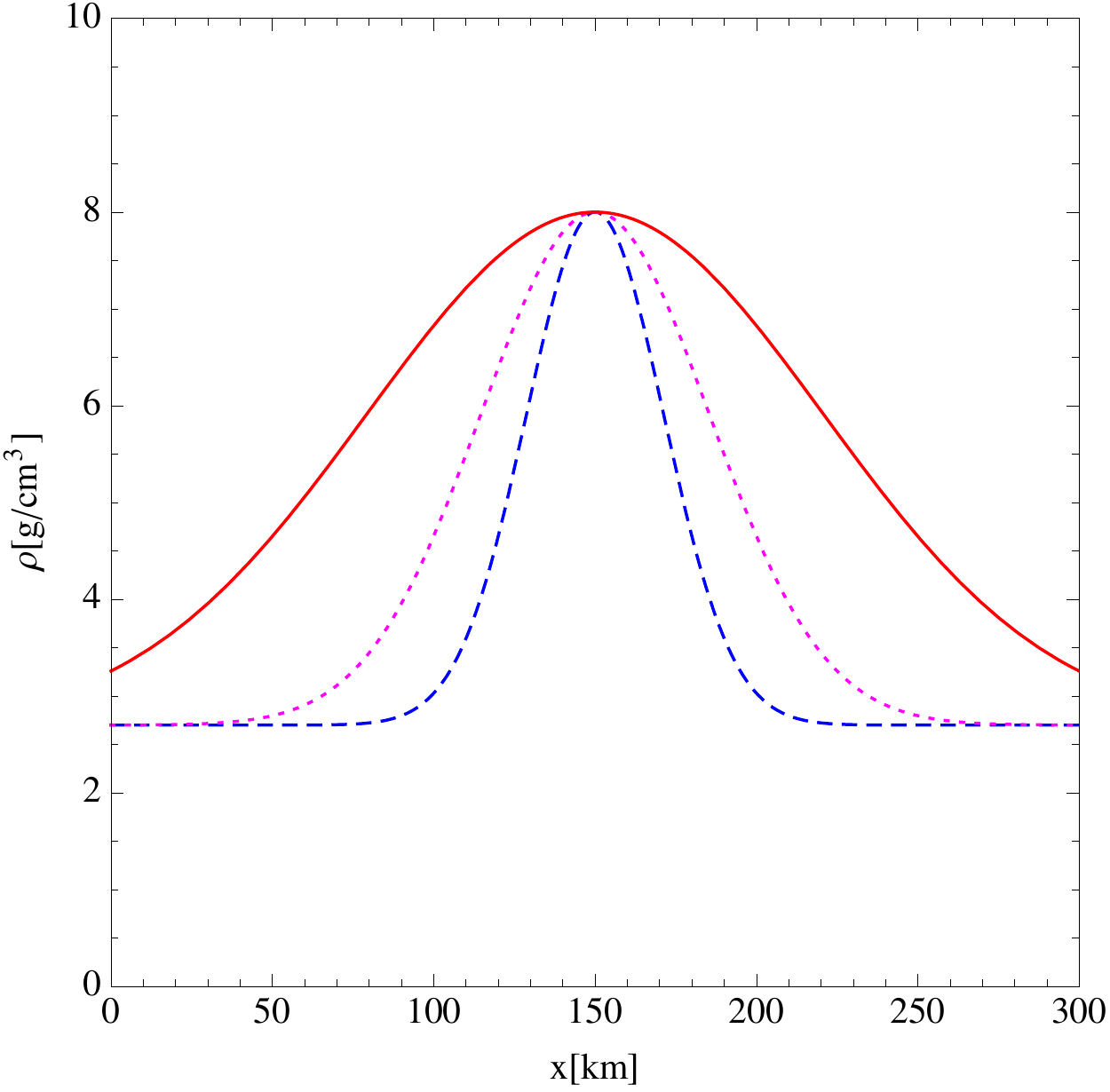}%
    \includegraphics[height=8.7cm,pagebox=cropbox]{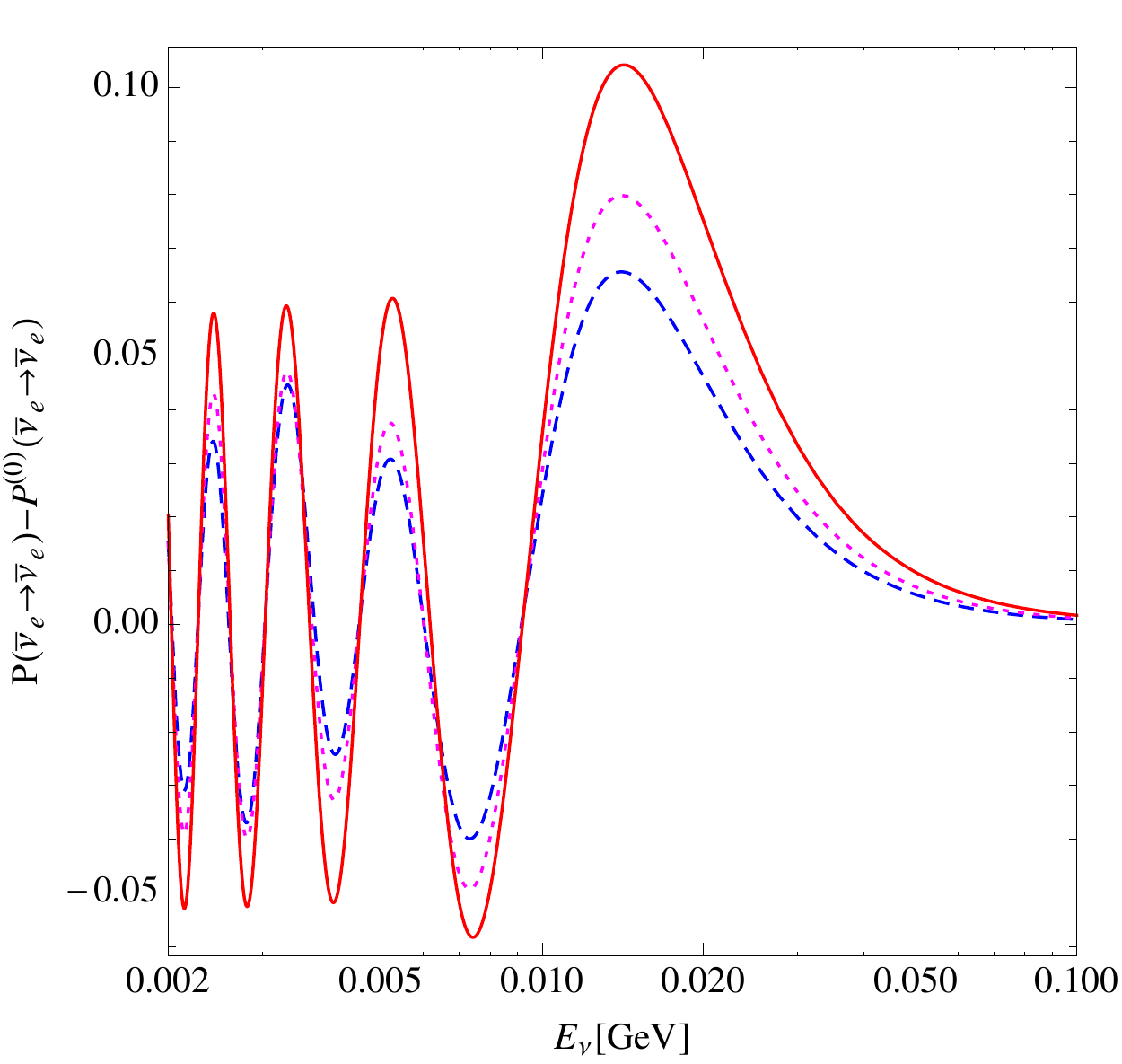}%
  }%
  \vspace{-2ex}
  \caption{
    Deviation of neutrino oscillation probability
    $P(\bar \nu_e \to \bar \nu_e)$ from the vacuum one (right panel) 
    due to the matter effect
    of the three different density profiles shown
    in the left panel.  The width of the lump is taken
    as $D_l = 100$~km (red solid line), $50$~km (magenta dotted line) and
    $30$~km (blue dashed line), respectively.
  }
  \label{fig:FIG_dp}
\end{figure}
We show in Fig.~\ref{fig:FIG_dp} the deviation of the oscillation
probability from the vacuum one as a function of neutrino energy when
the lump density is $\rho_l = 8.0$~g/cm$^3$ that is close to the iron
one.  It is seen that the deviation changes in accordance with the
density profile.  This correspondence is the essence of the
oscillation tomography.  Furthermore, the deviation due to the matter
effect is not so large and then the precise measurement of the energy
spectrum is a key for the realization of the tomography.

\section{Tomography by Neutrino Pair Beam}
The strong source of neutrino pairs consisting of 
$\nu_\alpha$ and $\bar \nu_\alpha$ ($\alpha = e, \mu, \tau$)
has been proposed recently~\cite{Yoshimura:2015rva,Yoshimura:2015ujr}.
The neutrino pairs are emitted from circulating heavy ions
which are in a quantum mixed state
$\cos \theta_c | g \rangle + e^{-i \epsilon_{eg} t} \sin \theta_c | e \rangle$
where the ground state of the ion $|g \rangle$ mixes with
its appropriate excited state $|e \rangle$ by the mixing angle
$\theta_c$.  The energy gap of these state is denoted by $\epsilon_{eg}$
and the coherence (mixture) is quantified by
$\rho_{eg} = \sin \frac{\theta_c}{2}$.
The properties of this neutrino pair beam are as follows:
the maximal value of the energy sum of neutrino and antineutrino
is $E_m = 2 \gamma \epsilon_{eg}$ where $\gamma$ is the Lorentz boost factor
of the circulated ions, the beam is well focused for the large $\gamma$ region
and the effective angular area is approximately given by $1/\gamma^2$,
and the spectrum of a neutrino (or an antineutrino) with energy $E_\nu$
is given by~\cite{Yoshimura:2015rva}
\begin{align}
  \frac{d\Gamma}{d E_{\nu}}
  &= 
  \frac{1}{21 \times 2^{10} \times \sqrt{6 \pi} \,  \pi^{4}} 
  a N |\rho_{eg}|^{2} \sqrt{\rho \, \epsilon_{eg}} 
  G_{F}^{2} E_{m}^{4} \frac{1}{\gamma} f(E_\nu/E_m) 
  \,,
\end{align}
where the function $f(x)$ is given by
\begin{align}
  f(x)
  & =
  \sqrt{x} \int_{0}^{1-x} dy y^{\frac{1}{2}} (y +x)^{\frac{5}{4}}
    (1-x-y)^{\frac{7}{4}} \,,
\end{align}
and $a$ represents factors of the transition dipole moment
and the number of neutrino generations.
In this analysis we consider the vector current contribution
of neutrino interaction with ionic electron since it gives
the oscillation behavior in the probability of the single
neutrino detection (while the other neutrino of the pair
is undetected)~\cite{Asaka:2015tal}.
$N$ and $\rho$ are the number and orbital radius of circulated ions.
Note that, when $x \ll 1$, the function $f(x)$ is approximately
given by
\begin{align}
  f(x)
  &\simeq 0.0494 \sqrt{x} \,.
\end{align}
As representative values we take here $a N |\rho_{eg}|^2 = 10^8$,
$\gamma = 10^4$, $\rho=4$~km and $\epsilon_{eg} = 50$~keV as in
Ref.~\cite{Yoshimura:2015rva}.  In this case the maximum energy is
$E_m = 1$~GeV and the total rate of the pair production is
estimated as $5.26 \times 10^{16}$~s$^{-1}$.

The high intensity $\bar \nu_e$ beam, called as the beta
beam~\cite{Zucchelli:2002sa}, has been proposed.  Its flux is much
larger than the current accelerator experiments as
$2.3$~cm$^{-2}$~s$^{-1}$.  On the other hand, that of the pair beam is
estimated as $4.65 \times 10^8$~cm$^{-2}$~s$^{-1}$, which clearly shows
that the precise measurement of the energy spectrum of the oscillation
probability is expected.

In this analysis we consider the detector with liquid argon of
fiducial volume $10^5$~m$^3$.
The signal event is given by the number of the $\bar \nu_e$ charged current interaction
$\bar \nu_e + p \to e^+ + n$.  The cross section is obtained at low energies
\begin{align}
  \sigma(E_{\nu})
  &= 
    \frac{|V_{ud}|^{2} E_{\nu}^{2} G_{F}^{2} m_{N}}
    {3 \pi (2 E_{\nu}+m_{N})^{3}} \,
    \Big[16 (g_{V}^{2} + g_{A} g_{V} + g_{A}^{2})
    E_{\nu}^{2} + 12 E_{\nu} m_{N} (g_{V}^{2} + g_{V} g_{A} + 2 g_{A}^{2}
    )+ 3 m_{N}^{2} (g_{V}^{2} + 3 g_{A}^{2})
    \Big]
    \nonumber \\
  & \simeq 9.34 \times 10^{-44} \; \mbox{cm}^{2}
    \left( \frac{E_{\nu}}{\mbox{MeV}} \right)^{2} ~~
    \mbox{for}~E_\nu \ll m_N\,,
\end{align}
where $m_N$ is a nucleon mass ($m_N=m_p$), $g_V = 1$ and
$g_A = 1.2695$.  The number of the signal events $N (E_\nu)$ for the
energy region $[E_\nu, E_\nu+\Delta E_\nu]$ in the detector located at
$L$ from the source is given by
\begin{align}
  \label{eq:Nsig}
   N (E_\nu)
   \simeq
   \frac{d\Gamma}{d E_{\nu}} \, \Delta E_{\nu} \,
   \frac{\gamma^{2}}{4 \pi L^{2}} \, P(E_{\nu},L) \,
   \sigma(E_{\nu}) \, n_{N} \, V_{d} \, T \,,
\end{align}
where $V_d$ and $n_N$ is the volume and the nucleon density of the
detector%
\footnote{
  For the detector of liquid argon
$\rho = 1.4$ g/cm$^3$ and 
$n_N = N_A \rho/(40 m_p) = 0.035 N_A$~cm$^{-3}$ where
$N_A$ is the Avogadro constant.}
and $T$ is the duration of the observation.
Here the area of the detector is assumed to be smaller
than $4 \pi L^2/\gamma^2$.
It is then for $E_\nu \ll E_m$ that
\begin{align}
  N (E_{\nu})
  \simeq
  4.73 \times 10^7
  \times
  P( \bar \nu_e \to \bar \nu_e ; E_\nu )
  \left( \frac{E_\nu}{100~\mbox{MeV}} \right)^{\frac 52}
  \left( \frac{\Delta E_\nu}{1~\mbox{MeV}} \right)
  \left( \frac{300~\mbox{km}}{L} \right)^2
  \left( \frac{V_d}{10^5~\mbox{m}^3} \right)
  \left( \frac{T}{1~\mbox{year}} \right) \,.
\end{align}
Here we have used the facts that, although the beam 
produces the pairs of neutrino and antineutrino with all flavors
through the charged and neutral current interactions,
the dominant one is the pairs of $\nu_e$ and $\bar \nu_e$, and that
the detection probability of $\bar \nu_e$ (while the other $\nu_e$ of the pair
is undetected) is approximately given by
the neutrino oscillation probability
$P( \bar \nu_e \to \bar \nu_e ; E_\nu)$~\cite{Asaka:2015tal}.
It is seen that a large number of $\bar \nu_e$ events can be
detected even at $L=300$~km.%
\footnote{
  The number of $\bar \nu_e$ events in the
  energy region $E_\nu = 2.6$--$8.5$~MeV for one year at KamLAND
  is ${\cal O}(10^2)$~\cite{Gando:2013nba},
  and that for the present case is
  $2.99 \times 10^6$.
}

In order to show the significance of the use of the neutrino pair beam
quantitatively, we perform the $\chi^2$ analysis
and show how precise the width and density of the lump can be
reconstructed when the density profile is given by Eq.~(\ref{eq:rho_ex}).
We consider the case when the energy spectrum is measured
for 100~bins ($N_b = 100$) from $E_\nu = 2$~MeV to 100~MeV,%
\footnote{
  The neutrino energy threshold of $\bar \nu_e + p \to e^+ + n$ is
  $E_* = [(m_n + m_e)^2 - m_p^2]/(2 m_p) \simeq 1.806$~MeV~\cite{Giunti_Kim}.
}
and $\Delta \chi^2$ is introduced by 
\begin{align}
  \Delta \chi^{2}
  = \sum_{i=1}^{N_b}
  \frac{ \left[ N(E_{i})|_{D_{\ast}, \rho_{\ast}}
  - N(E_{i})|_{D_{l}, \rho_{l}} \right]^{2} }
  {\sigma^2 (E_{i})} \, ,
\end{align}
where the true values of width and density of the lump are taken as
$D_l = 30$~km and $\rho_l = 8.0$~g/cm$^3$.
Here we take into account only the statistical error and
$\sigma(E_{i})= \sqrt{N(E_{i})|_{D_{l},\rho_{l}}}$.

\begin{figure}[t]
  \centerline{
   \includegraphics[width=13cm]{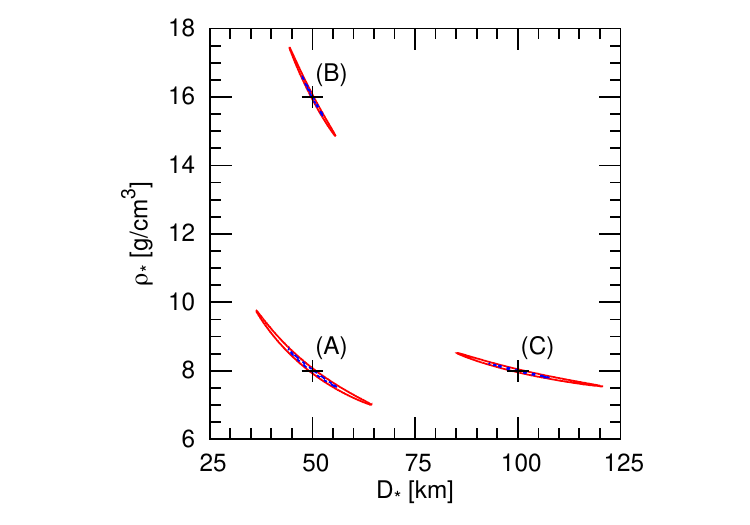}
  }%
  \vspace{-2ex}
  \caption{ Contour plot of $\Delta \chi^2$ in the $D_*$-$\rho_*$
    plane.  The true values of the width and density of the lump are
    taken as (A) $D_l=50$~km and $\rho_l=8.0$~g/cm$^3$,
    (B) $D_l=50$~km and $\rho_l=16$~g/cm$^3$,
    and (C) $D_l=100$~km and $\rho_l=8.0$~g/cm$^3$, respectively.
    $\Delta \chi^2 = 2.3$  ($1\sigma$ level) is shown
    by blue dashed lines and $\Delta \chi^2 = 11.83$  ($3\sigma$ level)
    is shown by red solid lines.}
  \label{fig:fig_contour}
\end{figure}
In Fig.~\ref{fig:fig_contour} contour plots of
$\Delta \chi^2$ in the $D_*$-$\rho_*$ plane are presented for three different cases.
The lines with $\Delta \chi^2 = 2.3$  ($1\sigma$ level)
and $\Delta \chi^2 = 11.83$  ($3\sigma$ level)~\cite{Patrignani:2016xqp}
are shown.
It is seen that there is an approximate degeneracy between $D_\ast$ and
$\rho_*$.  This is because the leading contribution to the
oscillation probability from the lump is proportional to the
combination $\Delta \rho_* \, D_\ast$ where
$\Delta \rho_\ast = \rho_\ast - \bar \rho$.
This degeneracy is broken by the higher order corrections of $\Delta \rho_*$ and $D_\ast$.
The pair beam can probe the lump at the 1~$\sigma$ level as
\begin{align}
  \begin{array}{l l c l}
    \mbox{(A)}
    &
      D_\ast = 50^{+5.9}_{-5.9}~\mbox{km}
    &
      \mbox{and}
    &
      \rho_\ast = 8.0^{+0.62}_{-0.48}~\mbox{g}~\mbox{cm}^{-3} \,,
    \\
    \mbox{(B)}
    &
      D_\ast = 50^{+2.5}_{-2.4}~\mbox{km}
    &
      \mbox{and}
    &
      \rho_\ast = 16^{+0.58}_{-0.53}~\mbox{g}~\mbox{cm}^{-3} \,,
    \\
    \mbox{(C)}
    &
      D_\ast = 100^{+8.2}_{-7.1}~\mbox{km}
    &
      \mbox{and}
    &
      \rho_\ast = 8.0^{+0.22}_{-0.21}~\mbox{g}~\mbox{cm}^{-3} \,,
  \end{array}
\end{align}

It is, therefore, found that the neutrino pair beam provides the powerful
tool to realize the oscillation tomography.
The density and width of the lump can be reconstructed accurately
if one takes into account the statistical error only.
In order to obtain the realistic precision of the
reconstruction we have to include various systematic errors in
production, propagation and detection rates as well as the mass squared
difference and mixing matrix of neutrinos.  This issue is beyond the
scope of this analysis.

\section{Reconstruction of Density Profile}
\label{sec:reconstruction}
Next, we turn to discuss the reconstruction procedure of $\rho (x)$.
One of the serious problems to perform the reconstruction is the
degeneracy of the oscillation probabilities.  In two-flavor neutrino
case the probability between flavor neutrinos with the density profile
$\rho(x)$ is the same as that with $\rho (L-x)$.  This means that the
measurement of the energy spectrum of the oscillation probability is
insufficient for the reconstruction when the density profile is
asymmetric, {\it i.e.}\,, $\rho(x) \neq \rho(L-x)$.  The
reconstruction without ambiguity is possible only for the symmetric
density profile with $\rho (x) = \rho(L-x)$.  It has been pointed
out~\cite{Akhmedov:2004rq,Akhmedov:2005yt} that this difficulty is
absent when one uses the solar and supernova neutrinos to probe the
interior of the Earth.  This is because the initial neutrinos before
entering the Earth are not in the flavor eigenstate but in the mass
eigenstate due to the MSW effect.

The tomography by the neutrino pair beam under consideration relies on
the oscillation probability of $\bar \nu_e \to \bar \nu_e$ as
explained in the previous section, and hence we are faced with this
problem.  In this analysis we merely assume the symmetric profile to
avoid it, which is the first step towards the analyses of more
complicated profiles.

Another difficulty of the reconstruction is the cost of numerical
calculations.  To reduce it we would like to propose a method based on
the perturbation of matter effects. Other methods will be discussed
elsewhere~\cite{Asaka:2018xxx}.  Notice that the matter effects can be
treated perturbatively for~\cite{Ioannisian:2004jk}
\begin{align}
  \label{eq:perturbation}
  \frac{\Delta m^2}{2 E_\nu} > \frac{G_F }{\sqrt{2} \,m_p} \, \rho \,.
\end{align}
Thus, the matter density and the neutrino energy should be sufficiently
small for a given $\Delta m^2$.

Our proposed method for the reconstruction is as follows: (i) First,
we discretize the neutrino baseline into the $N_L$ segments.
We take here $N_L=60$ as an example.  The matter densities at each
segment $\rho_i = \rho (x_i)$ ($i=1,2, \cdots, N_L$) are taken as free
parameters, which will be determined by using the $\chi^2$ fit to the
energy spectrum of the oscillation probability.

(ii) Second, we also divide the energy range of interest into the
$N_E$ parts.  Here the minimum energy is taken as 2~MeV, which is
larger than the threshold energy of $\bar \nu_e$ for the charged
current interaction (see the discussion in the previous section).  On
the other hand, the maximum energy is taken as 100~MeV 
which is sufficiently small to justify the perturbative
treatment of the matter effects. (See Eq.~(\ref{eq:perturbation})).
Here we divide into the $N_E = 100$ parts of even intervals.

(iii) Finally, we determine the densities $\rho_i$
by minimizing the $\chi^2$ function which compares the experimental data
$N^{\rm obs} (E_i)$ for a given density profile $\rho(x)$
with the theoretical predictions $N^{\rm th} (E_i)$
from unknown $\rho_i$.
The explicit form of the $\chi^2$ function is 
\begin{align}
  \chi^2 = \sum_{i=1, N_E}
  \frac{ \big[ N^{\rm obs} (E_i) - N^{\rm th} (E_i) \big]^2}
  {\sigma^2 (E_i)} \,,
\end{align}
where $\sigma (E_i) = \sqrt{N^{\rm obs} (E_i)}$.
In this analysis we take the number of signal events in Eq.~(\ref{eq:Nsig})
estimated from the oscillation probability including the
precise matter effect as the observation one $N^{\rm obs} (E_i)$.
On the other hand, the theoretical one $N^{\rm th} (E_i)$
is estimated from the analytical expression of the oscillation probability
obtained by the perturbation.  The use of the perturbative expression
is crucial in reducing the numerical cost to reconstruct $\rho_i$.

\begin{figure}[t]
  \centerline{
\includegraphics[height=8.5cm]{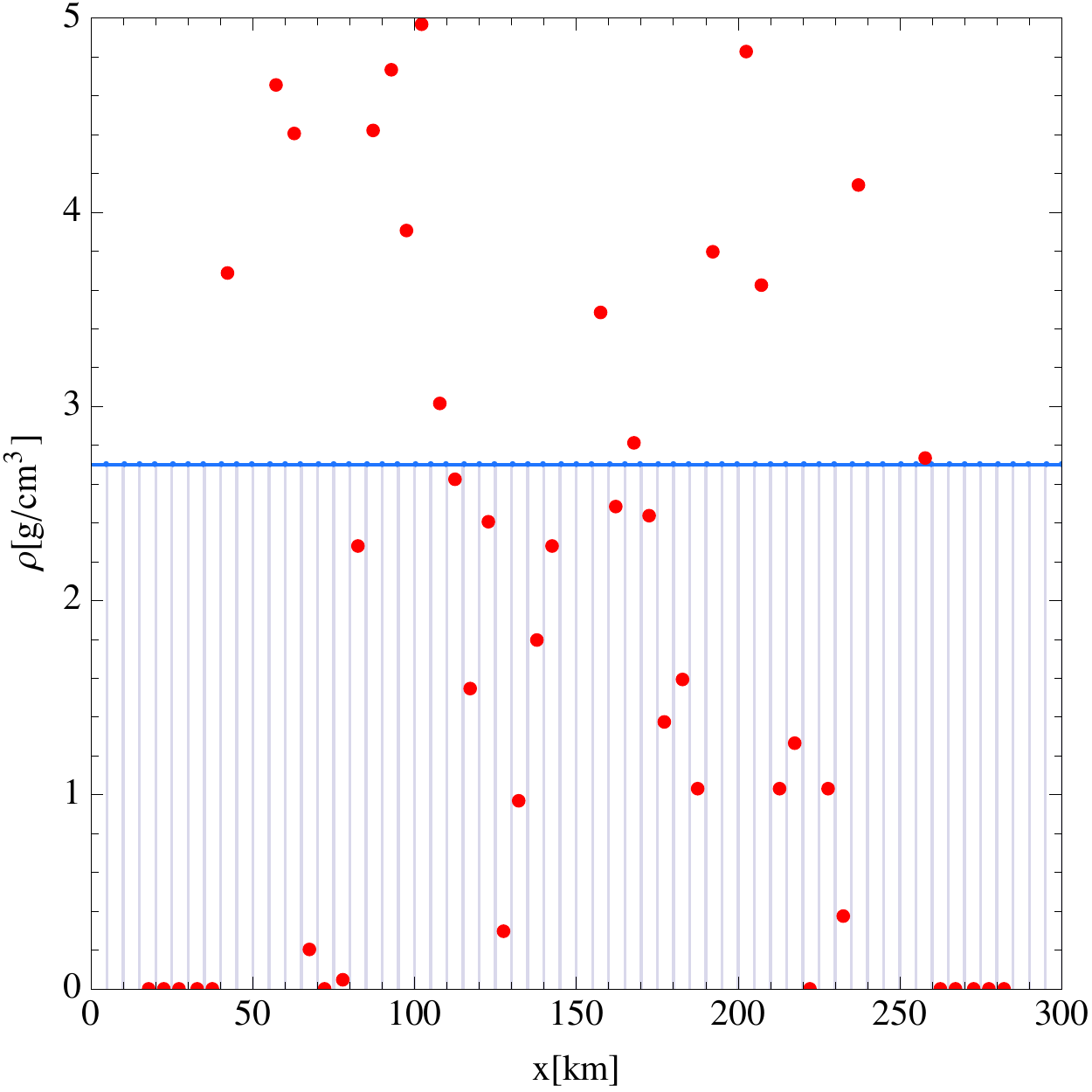}%
\includegraphics[height=8.5cm]{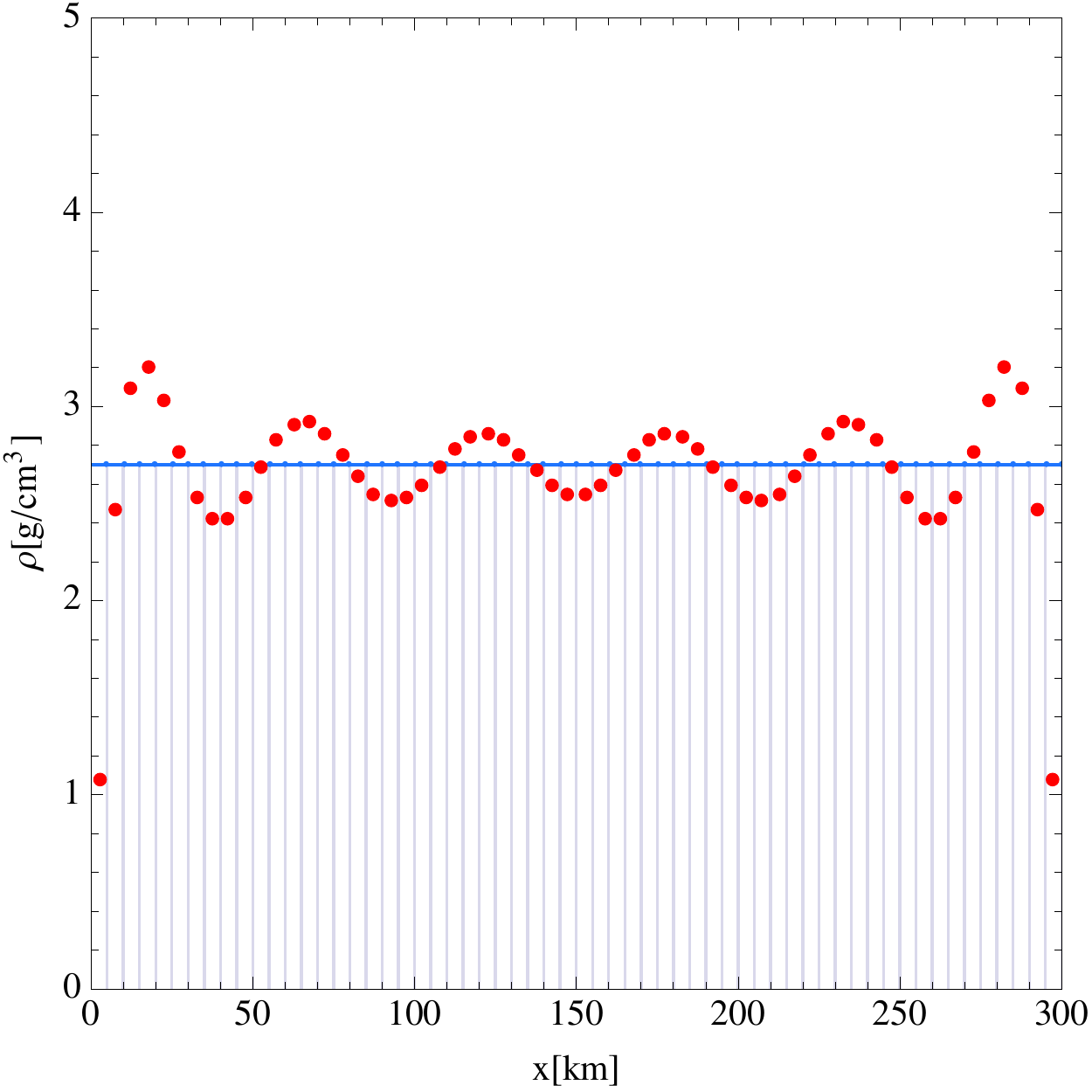}
  }%
  \vspace{-2ex}
  \caption{
    Results of the reconstruction of the flat density profile
    with $\rho(x) = \bar \rho = 2.7$~g/cm$^3$.  
    We use the analytic expressions of
    the energy spectra at
    the first order (left panel) and the second order (right panel), respectively.
    The true density profiles are shown by blue lines,
    while the reconstructed ones are shown by red points.
  }
  \label{fig:FIG_RC_rho_1st.pdf}
\end{figure}
First of all, we show the results for the case with the flat
density profile
\begin{align}
  \rho (x) = \bar \rho = 2.7~\mbox{g/cm}^3 \,.  
\end{align}
As shown in Fig.~\ref{fig:FIG_RC_rho_1st.pdf},
the reconstruction by using the probability
with the first order correction
$P^{(1)}(\bar \nu_e \to \bar \nu_e; E_\nu)$ is found to be unsuccessful
in the present case.
On the other hand,
when we include the second order correction
$P^{(2)}(\bar \nu_e \to \bar \nu_e; E_\nu)$,
the density profile can be reconstructed within a certain error.
It is found that the precision of the reconstruction
is ${\cal O}$(10)~\%
except for the regions near the production and detection points.

\begin{figure}[t]
  \centerline{
\includegraphics[height=8.5cm]{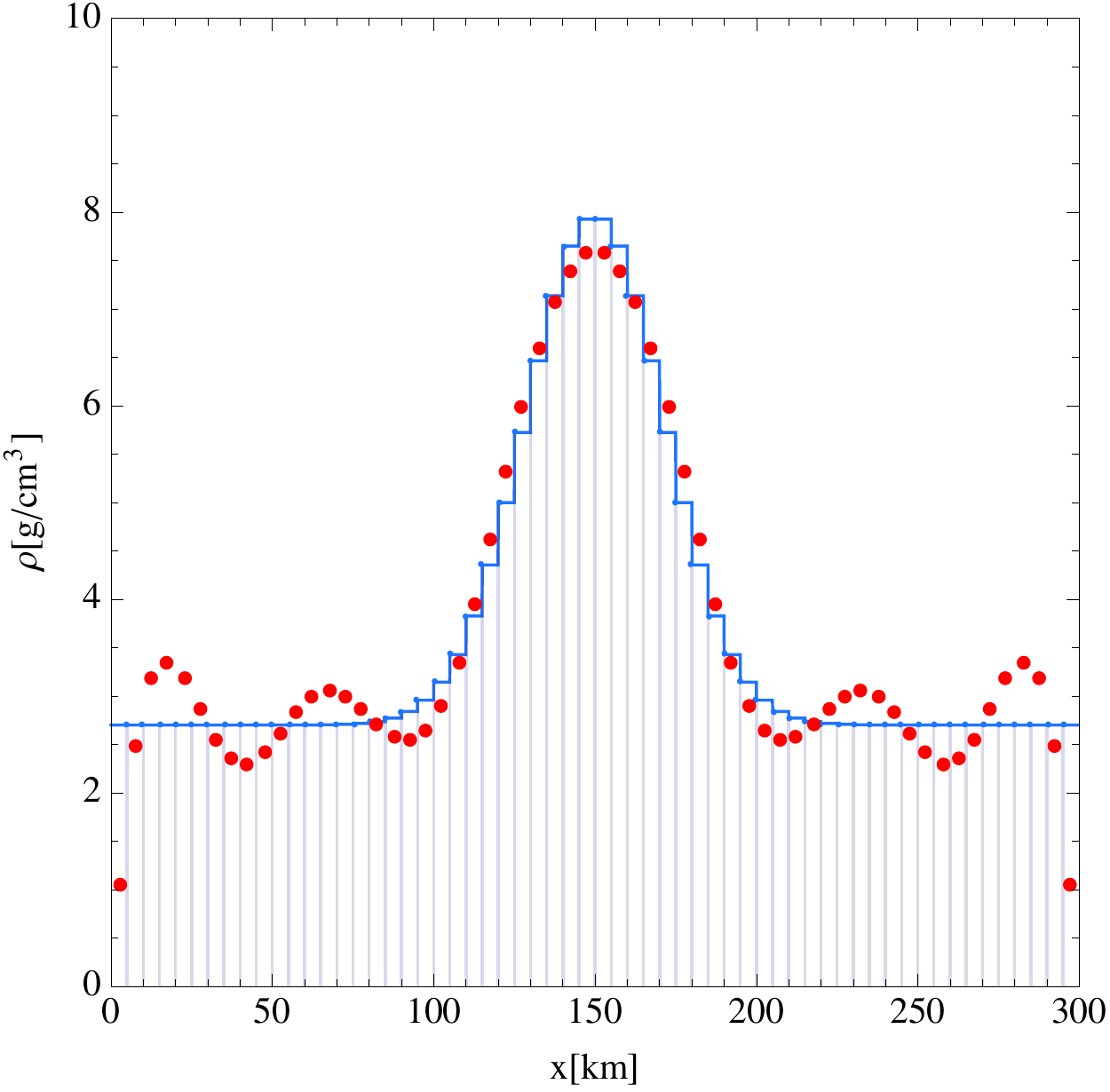}
\includegraphics[height=8.5cm]{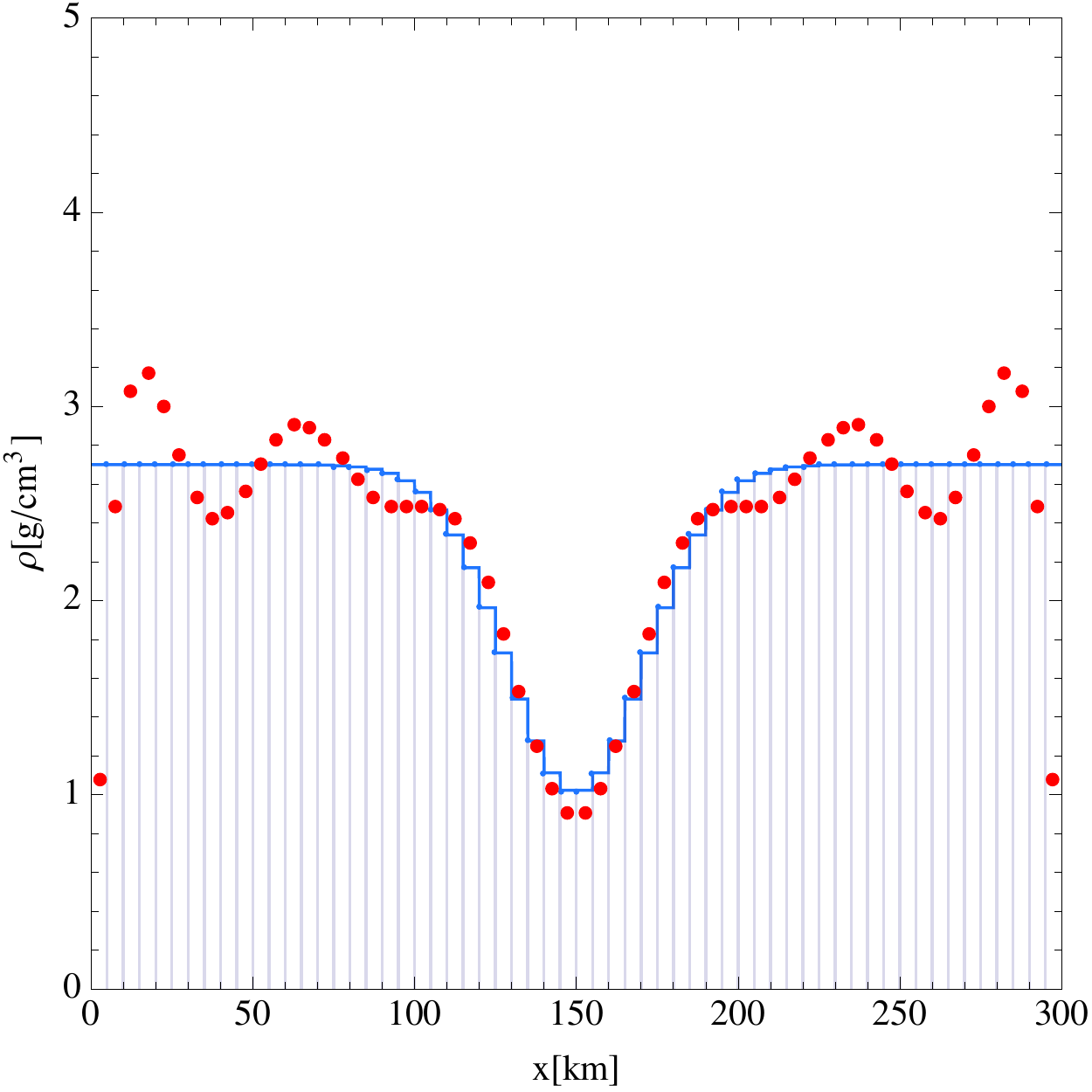}%
  }%
  \vspace{-2ex}
  \caption{
    Results of the reconstruction of the two density profiles.
    The density at the center is $\rho = 8$~g/cm$^3$
    that is for iron (left panel) and 1~g/cm$^3$ for water (right panel).
    The true density profiles are shown by blue lines,
    while the reconstructed ones are shown by red points.
  }
  \label{fig:iron_water}
\end{figure}
Next, we perform the reconstruction of the
density profile with a bump or a dip
by using the second order expression.
The results are shown in Fig.~\ref{fig:iron_water}.
The density at the center for the bump is taken as
8~g/cm$^3$ which is close to the value for iron, and
that for the dip is 1~g/cm$^3$ which is for water
(with a pressure of one atmosphere).  It is seen that
the profiles can be reconstructed well except for the
regions near the production and detection points.

\begin{figure}[t]
  \centerline{
   \includegraphics[height=8.5cm]{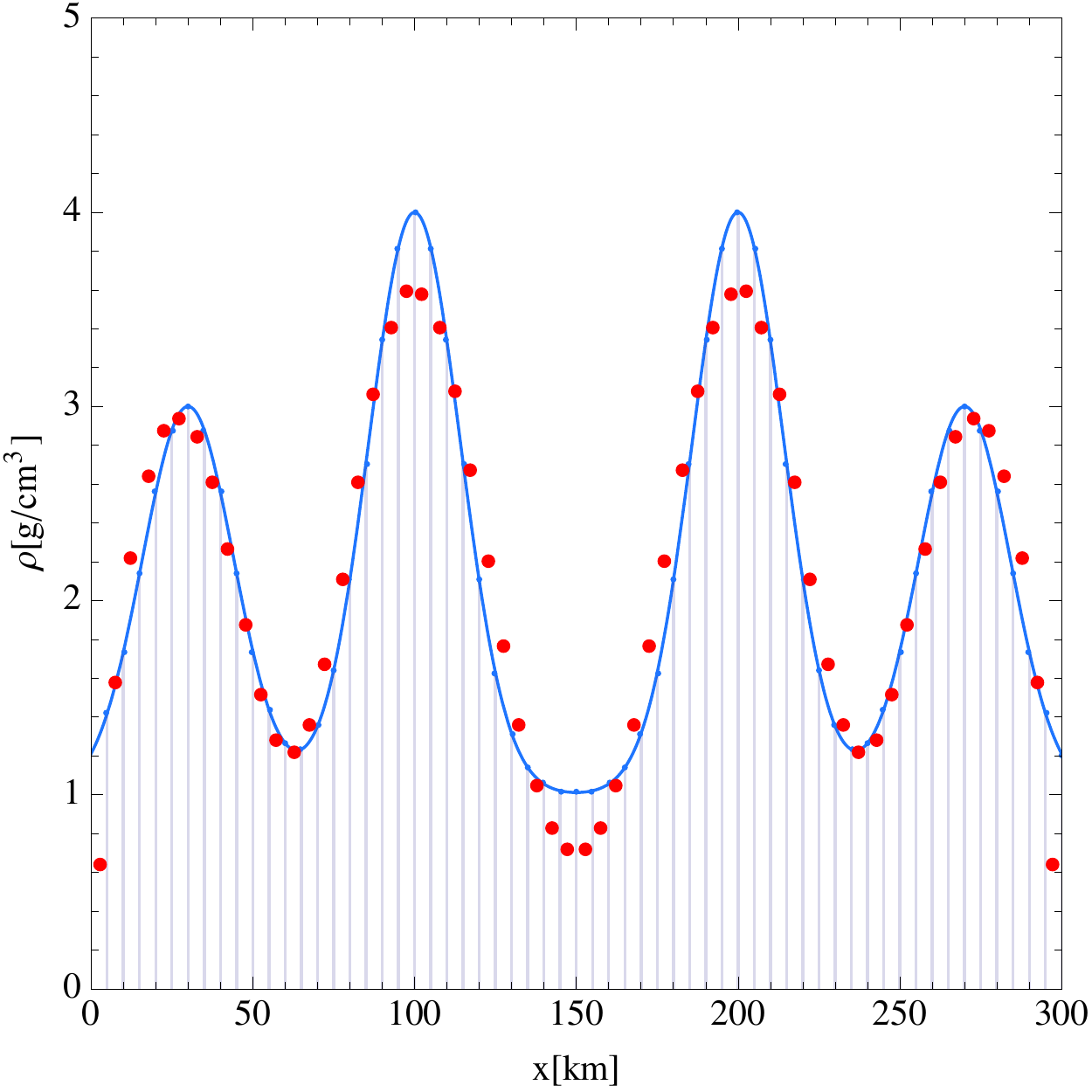}
  }%
  \vspace{-2ex}
  \caption{
    Result of the reconstruction of the symmetric density profile.
    The true density profile is shown by blue line,
    while the reconstructed one is shown by red points.
  }
  \label{fig:fig_fit_4lump}
\end{figure}
We have found that our proposed method at the second order perturbation
works successfully.  A rather complicated profile can be reconstructed
as long as the density profile is symmetric $\rho(x) = \rho(L-x)$.
See, for example, Fig.~\ref{fig:fig_fit_4lump}.  It should be
stressed that this method can operate even with the limited numerical
cost because of the analytical expression of the oscillation probability.

Finally, we have observed that the minimum energy in the oscillation
probability used for the reconstruction is important to obtain a better
result.  In the present analysis we take 2~MeV because of the
threshold energy of the $\bar \nu_e$ detection process.  As the minimum
energy becomes smaller, the result of the reconstruction becomes
better.  We have found in such cases that the reconstructed densities
even near the production and detection points become close to the true
values.  In addition, the oscillatory behavior of the reconstructed
profile at the flat density region becomes changed so that both
amplitude (deviation) and wavelength (interval) become
smaller, which leads to the better fit.  In other words, the main
source of the gap from the true profile might be the fact that we can
use the oscillation probability for the limited energy range.
Especially, those at low energies give a precise information of
the density profile.  Moreover, the corrections of matter effect
at higher orders are also the source of errors in the reconstruction.

\section{Conclusions}
We have investigated the oscillation tomography using the neutrino
pair beam.  It has been shown that the beam can offer the powerful
neutrino source to probe the Earth's interior and the precision
of the reconstruction of the density profile becomes improved
considerably.  In addition, we have proposed the tomography method
based on the perturbation of matter effects in the oscillation
probabilities.  This method works only for the symmetric profile
with $\rho(x) = \rho (L-x)$ under the situation we have discussed.
It has been demonstrated that 
the profile can be reconstructed well by including the second order
correction.  We believe that these two ingredients give considerable
progress toward the realization of the neutrino tomography.

\section*{Acknowledgments}
The work was partially supported by JSPS KAKENHI Grant Numbers
17K05410 (TA), 17H05198 (TA), 16H03993 (MT), 17H02895 (MT and MY),
17H05405 (MT), and 18K03621 (MT).
TA and HO thank the Yukawa Institute for
Theoretical Physics at Kyoto University for the useful discussions
during "the 22th Niigata-Yamagata joint school" (YITP-S-17-03).



\end{document}